\documentclass[12pt]{article}

\usepackage{arxiv}

\usepackage{amssymb}
\usepackage[romanian, english]{babel}
\usepackage{combelow}
\usepackage{newunicodechar}
\usepackage{graphicx}
\usepackage{mathpazo} 
\usepackage{microtype}
\usepackage{amsmath}
\usepackage{hyperref}
\usepackage{color}
\usepackage{url}
\usepackage{caption}
\usepackage[labelfont=bf]{caption}

\usepackage{graphicx}
\usepackage{dcolumn}
\usepackage{bm}
\usepackage[mathlines]{lineno}
\usepackage{color}
\usepackage{todonotes}
\usepackage{amssymb}
\usepackage{amsmath}
\usepackage{upgreek}

 \usepackage{floatrow}
\usepackage{graphicx}
\usepackage{subcaption}

\newunicodechar{Ș}{\cb{S}}
\newunicodechar{ș}{\cb{s}}
\newunicodechar{Ț}{\cb{T}}
\newunicodechar{ț}{\cb{t}}

\usepackage{authblk}

\renewcommand{\headeright}{}
\renewcommand{\undertitle}{}

\title{Multiple scattering camouflaged as magnetic stripes in single crystals of superconducting (La,Sr)$_2$CuO$_4$}

\author[1,2*]{A.-E. \c{T}u\c{t}ueanu} 
\author[1,2]{T.B. Tejsner}
\author[3,2]{M.E. L\v{a}c\v{a}tu\cb{s}u}
\author[1,2,4]{H.W. Hansen}
\author[4,2]{K.L. Eliasen}
\author[1]{M. Boehm}
\author[1]{P. Steffens}
\author[5]{C. Niedermayer}
\author[2]{K. Lefmann}

\affil[1]{Institute Max von Laue Paul Langevin, 38042 Grenoble, France}
\affil[2]{Nanoscience Center, Niels Bohr Institute, University of Copenhagen, 2100 Copenhagen {\O}, Denmark}
\affil[3]{Department of Energy Conversion and Storage, Technical University of Denmark, 2800 Kgs. Lyngby, Denmark}
\affil[4]{Glass and Time, IMFUFA, Department of Science and Environment, Roskilde University, 4000 Roskilde, Denmark}
\affil[5]{Laboratory for Neutron Scattering and Imaging, Paul Scherrer Institute, 5232 Villigen, Switzerland}
\affil[*]{email: tutueanu@ill.fr}

\date{}

\begin{document}
\maketitle

\begin{abstract}
Neutron diffraction has been a very prominent tool to investigate high-temperature superconductors, in particular through the discovery of an incommensurate magnetic signal known as stripes. We here report the findings of a neutron diffraction experiment on the superconductor (La,Sr)$_2$CuO$_4$, where a spurious signal appeared to be magnetic stripes. The signal strength was found to be strongly dependent on the neutron energy, peaking at $E = 4.6$~meV. We therefore attribute the origin of this signal to be a combination of multiple scattering and crystal twinning. A forward calculation of the scattering intensity including these two effects almost completely recovers our experimental observations. We emphasise the need for employing such analysis when searching for ways to avoid spurious scattering signals.
\end{abstract}


\section{Introduction}
\label{Introduction}

Since their discovery more than 30 years ago \cite{bednorz86}, considerable scientific effort has been put into the quest of understanding the emergence of superconductivity in ceramic cuprates. During this formidable amount of research, much evidence points to the interplay between magnetism and superconductivity as playing an important role for the mechanisms that form the superconducting Cooper pairs \cite{Fradkin15}.

Lanthanum-based cuprate superconductors possess a very complex electronic phase diagram with intertwined orders ranging from an antiferromagnetic Mott insulator to a metal. The long-range antiferromagnetic order in these compounds is destroyed by as little as $2\%$ strontium doping which introduces one electron hole per substituted atom \cite{dagotto1994correlated}. Upon hole doping, these materials are known to form so called magnetic stripes, also known as incommensurate antiferromagnetic (IC AFM) order, depicted as insulating domains of antiferromagnetically arranged spins separated by one dimensional "rivers of charge" within the CuO$_2$ layers.
Spin stripes were first proposed by Tranquada et al. in La$_{1.6-x}$Nd$_{0.4}$Sr$_x$CuO$_4$ \cite{tranquada1995evidence} and later confirmed in La$_{2-x}$Sr$_x$CuO$_4$ (LSCO), as summarized by Yamada et al. \cite{yamada1998doping}. In neutron scattering experiments, spin stripes are observed as a quartet of incommensurate peaks around the AFM reflection with a wavevector transfer of $\boldsymbol{q}_\textbf{IC} = (\pm\delta,1\pm\delta,0)$, 
or crystallographical equivalent positions, such as 
$\boldsymbol{q}_\textbf{IC} = (1\pm\delta,\pm\delta,0)$.

Since their initial discovery, numerous neutron scattering experiments have focused on characterizing the static and dynamic spin correlations under different conditions in the attempt of determining their relationship with superconductivity. The evidence so far points, in broad terms, to a competition between superconductivity and the static magnetic order, as the second was seen to vanish at optimal (and higher) doping \cite{julien2003magnetic} and is enhanced by an applied magnetic field, which is known to suppress superconductivity \cite{katano2000enhancement,lake2002antiferromagnetic}. 

From the early magnetic scattering studies performed on cuprates, multiple scattering was acknowledged as a possible contaminant of the stripe signal \cite{tranquada1995evidence}. Here we present the results of a systematic neutron study aimed at determining the magnitude and origin of the double scattering events in a highly underdoped superconducting La$_{1.93}$Sr$_{0.07}$CuO$_4$ single crystal. The findings presented here can be taken into account as guidelines in the planning phase of future experiments  (i.e. by fine tuning the neutron beam energy) in order to obtain optimal measurements of magnetic stripes.

\section{Experimental method} \label{experiment}

The sample used throughout this study is a highly underdoped La$_{2-x}$Sr$_x$CuO$_4$ crystal with nominal doping $x=0.07$. The large single crystal has been grown using the Traveling Solvent Floating Zone method and, later on, cut into two pieces, one annealed under oxygen gas flow as described in Ref.~\cite{fujita2002static} ($3.4$ g) which is the main sample of this paper and a second which was kept as-grown. The true (average) doping has been determined, before annealing, in a neutron diffraction experiment on the Morpheus diffractometer at the Paul Scherrer Institute (PSI), by following the temperature dependence of the strongly hole-doping dependent structural phase transition \cite{wakimoto2004neutron}, from the high temperature tetragonal phase (\textit{I4/mmm} space group number 139) to the low temperature othorombic one (\textit{Bmab} space group number 64). The measurements revealed a transition temperature $T_\text{s} = 362 \pm 1$ K which corresponds to a doping of $x=0.0753 \pm 0.0003$. The superconducting critical temperature of the as-grown and annealed crystals was established, through magnetic susceptibility measurements, to be $T_\text{c}^\text{onset} = 13.7 \pm 0.3$ K for both samples which is in agreement with measurements from literature on samples of similar doping \cite{kofu2009hidden}.

Elastic neutron scattering investigations were carried out on the cold-neutron triple axis spectrometer RITA II at PSI \cite{RITA2006}. The set-up contained a vertically focusing graphite monochromator, 80' horizontal collimation before the sample, a cooled beryllium filter after the sample in order to minimize the effects of higher order scattering contamination, followed by a graphite analyzer consisting of 9 blades, each 25~mm wide, and a coarse cross-talk-avoiding radial collimator before a position-sensitive detector. The sample was aligned with the CuO$_2$ layers in the scattering plane, making it possible to access wavevectors in the $(h,k,0)$- plane. 

Throughout this paper the orthorhombic notation is used, meaning that a commensurate antiferromagnetic reflection would be observed at $\boldsymbol{q} = (0,1,0)$, and crystallographically equivalent positions, and the stripe signal, due to its incommensurate magnetic modulation, should be visible at $(\pm \delta, 1\pm \delta, 0)$, and equivalent positions, where $\delta \sim x$ in this low-doping regime \cite{yamada1998doping}. In this notation, the 10~K lattice parameters of a $x=0.0750$ doped crystal (a value very similar to the doping level of our crystal) were earlier reported as $a_\text{o} = 5.33$ $\mathring{\text{A}}$, $b_\text{o} = 5.38$ $\mathring{\text{A}}$, $c_\text{o} = 13.16$ $\mathring{\text{A}}$ \cite{radaelli1994structural}.

To increase the efficiency of grid scanning around the antiferromagnetic reflection we made use of the monochromatic imaging mode of the multi-blade analyzer. In this configuration, all 9 blades elastically scatter neutrons of the same energy, but due to the difference in scattering angle, each blade maps a slightly different portion of reciprocal space \cite{Bahl2006}. 
In this configuration, the effective collimation is determined by the width of the sample, the (narrow) analyzer, and the distance between them. This leads to a collimation sequence of open -- 80' -- 40' -- open. However, the results presented throughout the paper are not particular to this set-up. The same observations can be reproduced on a standard triple axis spectrometer with a single-blade analyzer. 

\section{Results}
\label{Results}

\begin{figure}[h!]
\centering
\includegraphics[width=0.5\linewidth]{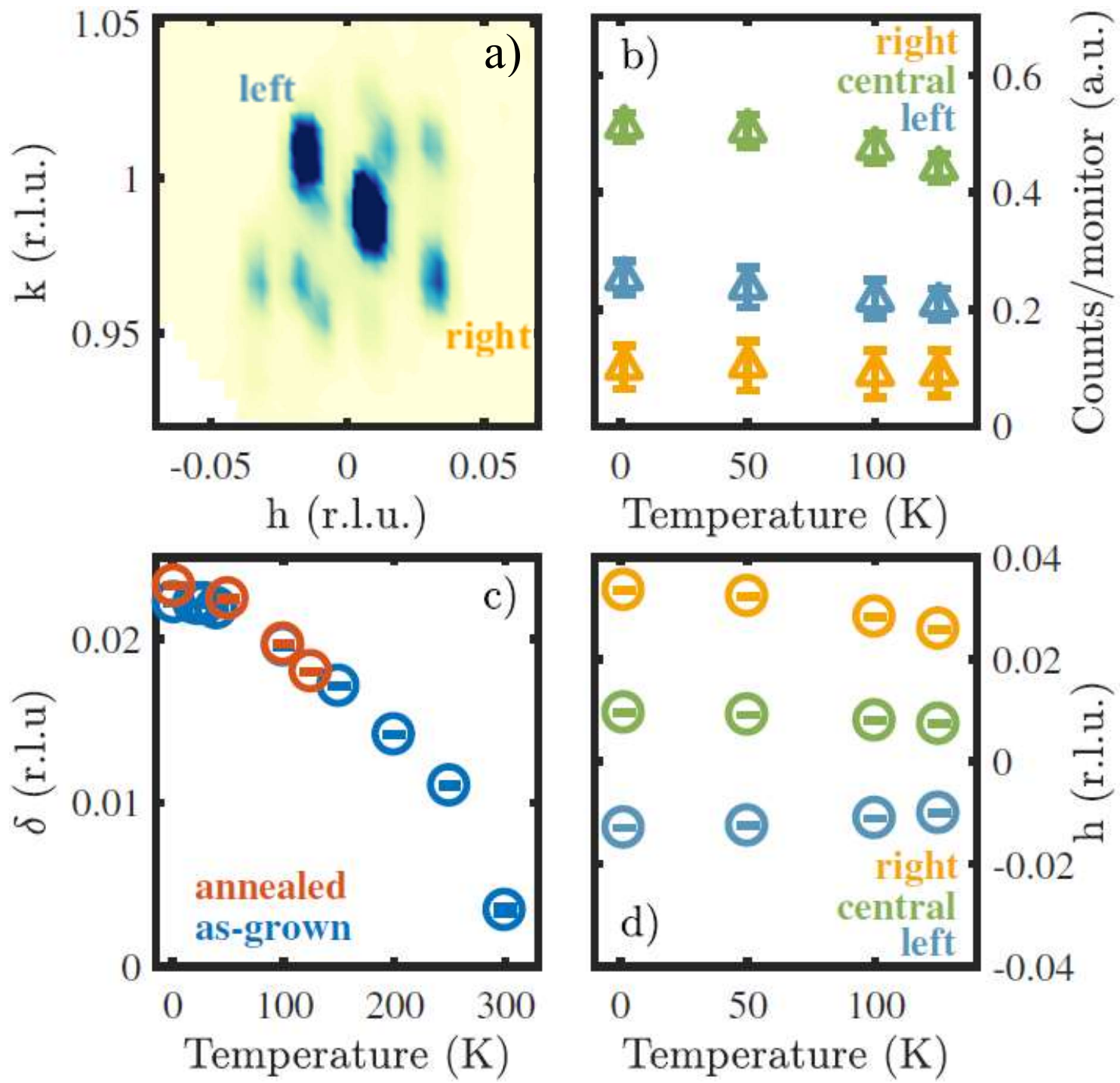} 
\caption{\label{fig:grid and temp dependence} a) Elastic scattering grid scan around the $(0,1,0)$ reflection taken at $2.5$ K on LSCO with $x = 0.07$ (annealed crystal). Temperature dependence of the b) peak intensity obtained from gaussian fits, c) incommensurability for both annealed and as-grown crystals of the same doping and d) peaks position. The incommensurability is defined as half the distance between the two side peaks marked as 'left' and 'right' in sub-figure a). }
\end{figure}

Initial neutron diffraction measurements were performed with incident and outgoing energies of $4.6$ meV, corresponding to $k_{\rm i} = k_{\rm f} = 1.49$ $\mathring{\text{A}}^{-1}$, where $k_{\rm i}$ and $k_{\rm f}$ are the wave vectors of the initial and final neutron beams, respectively. This is a standard configuration of cold triple-axis instruments that allows for an optimal filtering of higher order scattering by cooled polycrystalline beryllium.  
The measurements revealed the presence of a peak at the antiferromagnetic position, which has previously not, to our knowledge, been reported in homogeneously doped systems of $x>0.02$.  In addition to this central peak, a number of satellite peaks at incommensurate positions $(\pm \delta_\text{obs}, \, 1\pm \delta_\text{obs}, \, 0)$ were also measured, as depicted in Figure \ref{fig:grid and temp dependence}a, with incommensurability $\delta_\text{obs} = 0.023$, which is much smaller than the value expected from the dopant concentration: $\delta_x \sim x \sim 0.07$.

The temperature dependence of the signal, shown in Figure \ref{fig:grid and temp dependence} was, however, significantly different from that of static magnetic stripes. First of all, the intensity persisted up to room temperature with only a slight decrease as a function of increased temperature in contrast with genuine spin stripes signal that usually correlates with $T_\text{c}$ \cite{croft2014charge}. Secondly, a decrease in incommensurability with increasing temperature was measured (Figures \ref{fig:grid and temp dependence}c,d), whereas the spin modulation in stripe signals was earlier found to be invariable to temperature variations. These two temperature effects were invariant to sample annealing (Figure \ref{fig:grid and temp dependence}c).

Because the temperature dependence of the peaks resembles that of the tetragonal-to-orthorhombic phase transition around 360~K, multiple scattering involving reflections allowed only in the orthorhombic phase was soon proposed as the origin of the signal. 
Following the textbook by Shirane, Shapiro, and Tranquada \cite{shirane2002neutron}, the strict condition for multiple scattering to occur is that more than one reciprocal lattice point lies on the surface of the Ewald sphere. Its presence can be tested by either rotating the crystal around the contaminated scattering vector or by changing the neutron energy of the diffraction experiment ({\em i.e.}\ changing the radius of the Ewald sphere). It should be mentioned that in a real experiment, where the incident beam is not perfectly monochromatized, the Ewald sphere should be seen rather as a spherical shell, the thickness of which corresponds to the spread in wave vectors of the neutron beam.

\begin{figure}[h!]
\centering
    \includegraphics[width=0.6\linewidth]{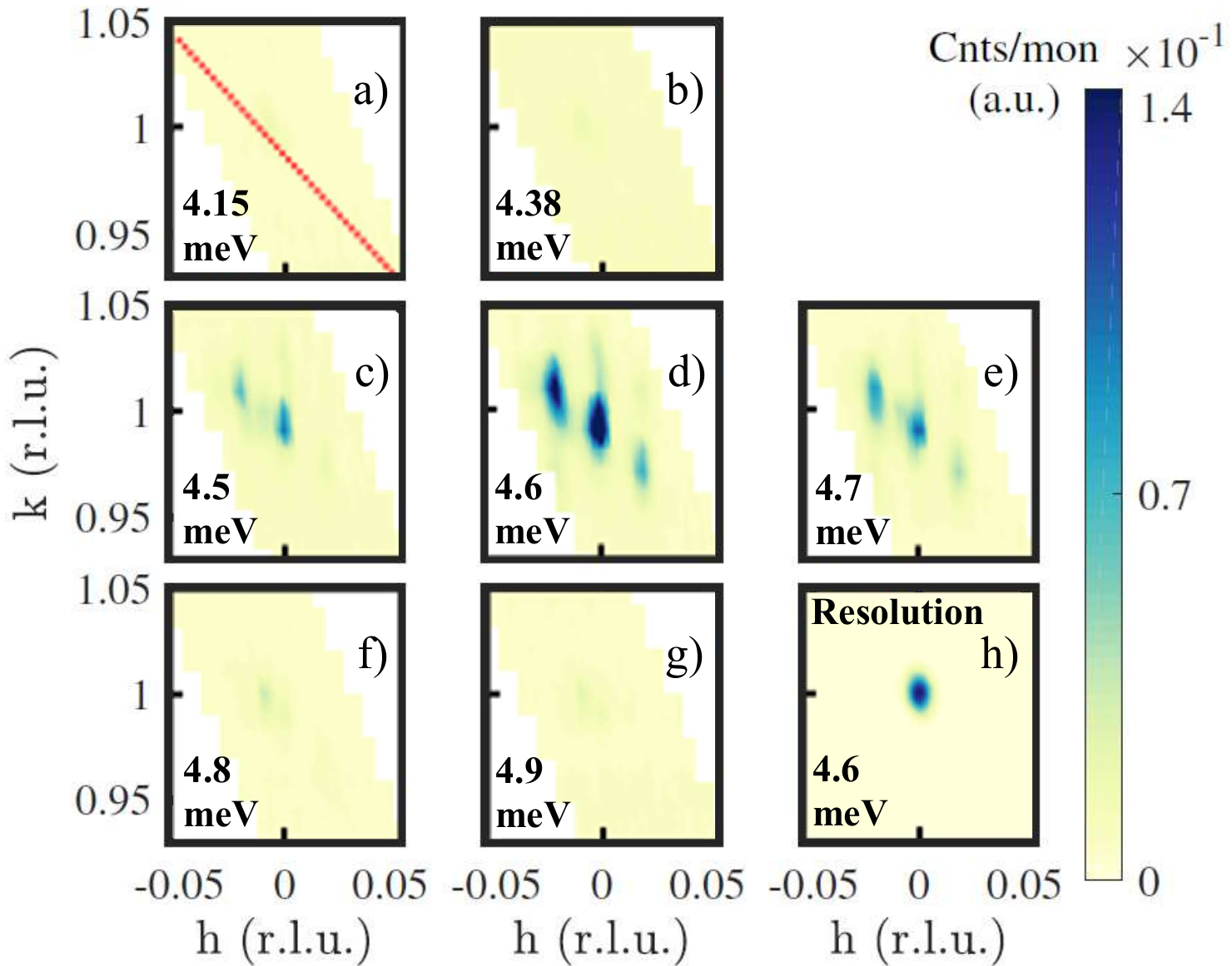}
\caption{\label{fig: grid energy}Diffraction measurements performed around the $(0,1,0)$ reflection at $125$ K. The caption in each figure indicates the value of the initial and final energy of the neutron beam. Sub-figure a) shows in red the direction of the two diagonal scans plotted in $2$D in Figure \ref{fig: grid cut}. Panel h) shows the simulated resolution of the instrument as presented in the text.}
\end{figure}

\begin{figure}[h!]
\centering
    \includegraphics[width=0.5\linewidth]{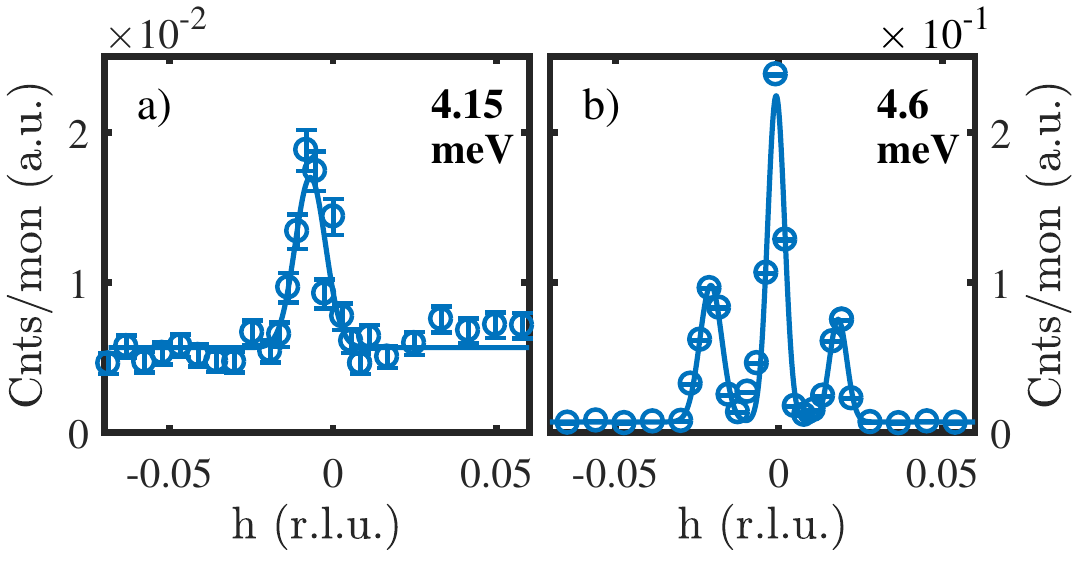}
    \caption{\label{fig: grid cut} Elastic diagonal scans taken at different values of the initial and final energy, noted in the caption, in the direction exemplified in Figure \ref{fig: grid energy}a. The solid line represents a gaussian fit to the data.}
\end{figure}

The method of changing wavelength of the diffraction experiment was employed in our experiments at RITA-II and the data is shown in Figure \ref{fig: grid energy}. Our findings clearly support the hypothesis of multiple scattering, since the incommensurate side peaks were only found at certain energies ($4.5$ meV to $4.7$ meV) and the intensity of the central $(0,1,0)$ peak greatly decreased when the initial and final energies were modified. 
The more than one order of magnitude difference in intensity can be visualised in Figure \ref{fig: grid cut} where $1$ dimensional cuts thought the grids are plotted for two different energies.
\par
Figure~\ref{fig: grid energy} also shows the result of a virtual experiment by the McStas ray-tracing simulation package \cite{lefmann2008,willendrup2019} of a hypothetical crystal showing single scattering at the (0 1 0) reflection, in effect providing the resolution function of the instrument. 
Through the same McStas simulation it was found that the width of incoming energy, $E_i$, on the sample was 0.13~meV (FWHM). This value is the relevant number to determine the effective thickness of the Ewald sphere and is not equal to the energy resolution of the spectrometer (which is 0.20~meV FWHM), because the energy resolution of the secondary spectrometer is here not taken into account.
\par

\section{Discussion}
\label{Discussion}
\subsection{Double scattering}
\begin{figure}[h!]
\centering
    \includegraphics[width=1\linewidth]{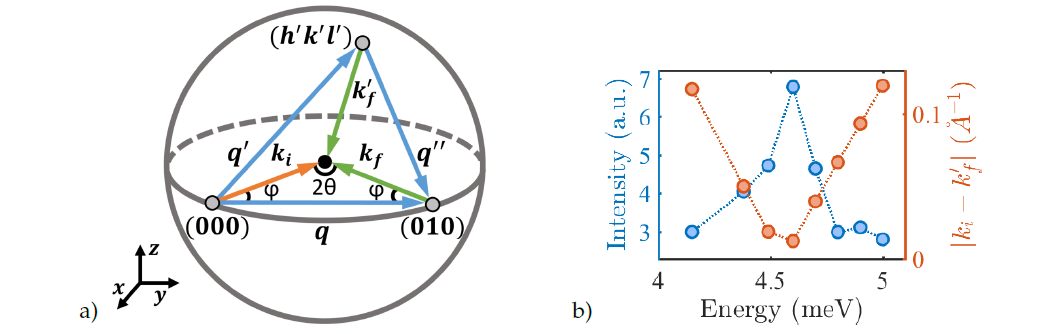}
    \caption{\label{fig:ewald} (a) Graphical representation of an instance of double scattering within the Ewald sphere. \textbf{$\boldsymbol{k_\text{i}}$} and \textbf{$\boldsymbol{k_\text{f}}$} define the incoming and outgoing neutron beam, $\boldsymbol{q^{\prime}}$ and $\boldsymbol{q^{\prime\prime}}$ denote the tested reflections allowed in the LTO phase and \textbf{$\boldsymbol{k_\text{f}^{\prime}}$} is the secondary reflection wavevector. (b) Blue symbols show the integrated intensity of the measured signal (see Figure \ref{fig: grid energy}) and follow the left side y-axis. The red symbols follow the right side y-axis and show the calculated difference between the magnitude of the two scattering wavevectors \textbf{$\boldsymbol{k_\text{i}}$} and \textbf{$\boldsymbol{k_\text{f}^{\prime}}$} for the case $\boldsymbol{q'} = (0,2,-2)$ and $\boldsymbol{q''} = (0,-1,2)$. The x-axis shows the incoming and outgoing energy of the neutron beam ($E_\text{i} = E_\text{f}$).}
\end{figure}

Due to the fact that the signal appears in a very narrow energy range and because we know the space groups of the crystallographic phases allowed in this sample, we were able to determine the reflections involved in a potential double scattering event. Following the treatment by Shirane, Shapiro, and Tranquada \cite{shirane2002neutron}, we construct the Ewald sphere corresponding to our experiment as shown in Figure \ref{fig:ewald}a. In our particular case, the condition for double scattering is fulfilled if another allowed reflection of the low temperature orthorombic (LTO) phase lies on the surface of the sphere. This poses two requirements: (1) the existence of a secondary reflection wavevector \textbf{$\boldsymbol{k_\text{f}^{\prime}}$} of the same length as \textbf{$\boldsymbol{k_\text{i}}$}; 
(2) the two scattering wavevectors  $\boldsymbol{q^{\prime}}$ and $\boldsymbol{q^{\prime\prime}}$ should correspond to allowed reflections in the LTO phase (space group no. $64$ in the unconventional \textit{Bmab} setting \cite{Bmab}). \textbf{$\boldsymbol{k_\text{f}^{\prime}}$} is geometrically defined as:

\begin{align}
     \boldsymbol{k_f^\prime} = \boldsymbol{k_i} -\boldsymbol{q^\prime}
\end{align} \
  
Using the scattering angle $\theta = \arcsin{(q/2k_i)}$ to calculate the angle $\varphi = 90-\theta$ and knowing the lattice parameters corresponding to the LTO phase we are able to rewrite the components of $\boldsymbol{k_\textbf{i}}$ and $\boldsymbol{q^{\prime}}$ as:

\begin{align}
q^{\prime}_x = h^{\prime} \cdot \frac{2\pi}{a_\text{o}}
\qquad 
q^{\prime}_y = k^{\prime} \cdot \frac{2\pi}{b_\text{o}}
\qquad 
q^{\prime}_z = l^{\prime} \cdot \frac{2\pi}{c_\text{o}}
\end{align}

\begin{align}
k_{\text{i}x} = k_\text{i} \sin{\varphi}
\qquad 
k_{\text{i}y} = k_\text{i} \cos{\varphi}
\qquad 
k_{\text{i}z} = 0
\end{align}\

where $h^{\prime}$, $k^{\prime}$ and $l^{\prime}$ are the Miller indices of the allowed reflections in the LTO phase. We have investigated indices with values between $-3$ and $3$. This decomposition allows us to calculate the corresponding $k_\text{f}^{\prime}$ in the energy range $4.5 - 4.6$ meV for all the allowed reflections of the LTO phase:

\begin{align}
    k_f^\prime = \sqrt{(k_{\text{i}x}-q_x^\prime)^2+(k_{\text{i}y}-q_y^\prime)^2+(k_{\text{i}z}-q_z^\prime)^2}
\end{align}\

Checking the above mentioned condition (1), meaning comparing the values of $k_\text{i}$ and $k_\text{f}^\prime$, two pair of reflections appear to lie on the surface of the Ewald sphere $(0,2,-2)$ (equivalent to $(0,2,2)$) and $(1,1,-3)$ (equivalent to $(1,1,3)$). However, condition (2) is only fulfilled for the first pair since both $(1,1,-3)$ and $(1,1,3)$ require the disallowed $\boldsymbol{q^{\prime\prime}}$ reflections ($(-1,0,3)$ and $(-1,0,-3)$ respectively) in order to produce intensity at $\boldsymbol{q} = \boldsymbol{q^{\prime}}+\boldsymbol{q^{\prime\prime}} = (0,1,0)$. In contrast, $(0,2,-2)$ and $(0,2,2)$ reflections pair with $\boldsymbol{q^{\prime\prime}} = (0,-1,2)$ and $(0,-1,-2)$ respectively, which are only allowed in the orthorombic phase and not in the tetragonal one. This explanation is thus in direct agreement with the specific temperature dependence of the signal as presented in Figure \ref{fig:grid and temp dependence}.

Figure \ref{fig:ewald}b shows the excellent correlation between the fulfillment of condition (1) for the $(0,2,-2)$ reflection and the peak in integrated intensity of the measured data. The red symbols close to $y=0$ correspond to the fulfillment of the first condition for multiple scattering: $k_f^{\prime}\simeq k_i$ for the $(0,2,-2)$ which takes place in the same energy range as the observed increase in scattered intensity. The $(0,2,2)$ reflection behaves in an equivalent manner.

\subsection{The effect of twinning}
We attribute the complex pattern of multiple peaks around the $(0,1,0)$ reflection, as it can be observed in Figure \ref{fig:grid and temp dependence}a, to a combination of the double scattering mechanism, presented above, and twinning of the reflections. It is well known that LSCO crystals in the low temperature orthorombic phase are composed of two sets of two twin orientations sharing the $(1,1,0)$ or $(1,-1,0)$ planes \cite{braden1992characterization}. This co-existence of a $4$-phase structure with equal lattice parameters but different orientations of the $(a,b)$ planes is observed in neutron scattering experiments as a peak
splitting of the Bragg reflections along the $h$ and $k$ directions (as exemplified in Figure 1 from Ref.~\cite{braden1992characterization}). \par
\begin{figure}[h!]
\centering
\centering  
\includegraphics[width=0.5\linewidth]{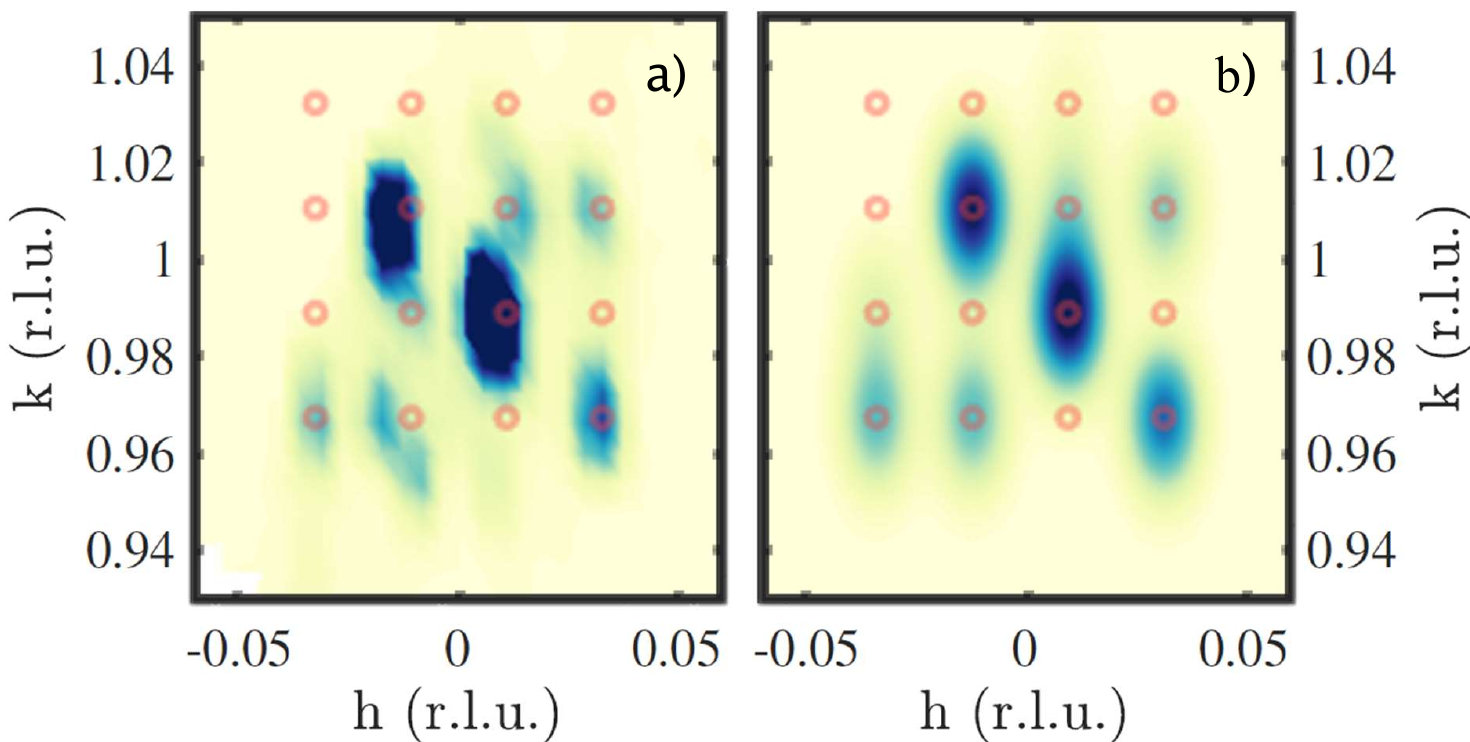} 
\caption{\label{fig:twinning} (a) Elastic scattering grid scan around the $(0,1,0)$ reflection taken at $2.5$ K on LSCO with $x = 0.07$ annealed. (b) Calculated intensity patterns of the 16 peaks originating from the twinning of the sample. The red circles indicate the position of all reflections. The intensity of each gaussian distribution around the calculated position is tuned as to match the measured data.}
\end{figure}

Thus, if we consider all possible combinations of $\boldsymbol{q} = \boldsymbol{q^{\prime}} + \boldsymbol{q^{\prime\prime}}$, where $\boldsymbol{q^{\prime}}$ and $\boldsymbol{q^{\prime\prime}}$ are the wavevectors of the $4$ satellite peaks around $(0,2,2)$ and $(0,-1,-2)$ respectively (i.e. $(\pm 2t,2\pm 2t,2)$ and $(\pm t,-1\pm t,-2)$), we obtain a total of $16$ peaks appearing around the $(0,1,0)$ reflection (as depicted by the red circles in Figure \ref{fig:twinning}). Note that the incommensurability ($t$) of the satellite peaks is related to the twin splitting ($\Delta = 90^\circ  - 2\arctan(b/a)$ \cite{braden1992characterization}) as $t=\Delta / 2$, in the small angle approximation, and it does not affect the $l$ value since there is no splitting along the c-axis. In Figure \ref{fig:twinning}b we plot the calculated pattern obtained as gaussian distributions centered at the calculated peak positions and with tuned amplitudes as to match the measured data shown in Figure \ref{fig:twinning}a. The twin splitting used in the calculations corresponds to a lattice parameters ratio $b/a = 1.022$, a value slightly higher than the one from the literature $b_\text{o}/a_\text{o} = 1.009$ \cite{radaelli1994structural}. As similar figure could likely be produced from a true simulation of the scattering intensities by modifying the volume fractions of the 4 twin phases. However, we have chosen not to address this involved optimisation scheme, since we do not think is would add to the general understanding. The good agreement between the measured data and the calculated reconstruction implies that the observed multiple peaks pattern can be regarded as a superposition of the twinning patterns of all the reflections that simultaneously intersect the surface of the Ewald sphere.\par

\subsection{Earlier studies of stripes in cuprates}

Although numerous studies from the literature follow the behaviour of static spin stripes in LSCO superconductors as a function of doping, temperature and applied magnetic field, this spurious process has not been addressed in any of the publications to our knowledge. There are two main ways in which the additional double scattering signal went unreported: a) at higher doping ($x \sim 0.12$) the true magnetic signal appears at higher $\boldsymbol{q}$ values compared to the spurious one, thus, depending on the scanning direction, one is able to measure the stripe signal without detecting the spurion (for example by performing scans of the sample orientation, A3). Secondly, as we have argued throughout this paper, by tuning the incoming and outgoing neutron beam energy the contamination can easily be avoided. Indeed, many of the experiments presented in the literature have been carried out at energies outside the $4.5 - 4.7$ meV regime \cite{lake2002antiferromagnetic,kofu2009hidden, romer2013glassy, chang2008tuning, khaykovich2005field}.
\par
That multiple scattering can make forbidden peaks appear was already mentioned by the first discovery of stripes \cite{tranquada1995evidence}, where Tranquada {\em et al.}\ reported a neutron energy (13.9~meV) of being almost free of spurious scattering at (0,1,0). In this respect, not all findings in the present article are new as such. However, the effect of multiple scattering, and in particular the effect that twinning causes IC peaks to appear, is in our work thoroughly documented, as a warning and explanation to future experimentalists.

\section{Conclusion}
\label{Conclusion}

Our neutron diffraction measurements and calculations support the hypothesis of a double scattering event, originating from the $(0,2,-2)$- and $(0,-1,2)$-type reflections, corroborated with twinning of the crystal, as the cause for the observed varying intensity of the $(0,1,0)$ reflection and the appearance of incommensurate peaks around it. 

Because the intensity of the static magnetic stripe signal is very low in cuprates in general and in LSCO in particular \cite{tutueanu2019}, the contribution of multiple scattering should always be carefully considered and the neutron energy should be tuned as to minimize this intrinsic effect, which could easily lead to detrimental pollution of the experimental data. In particular for cold-neutron experiment on the LSCO-family of compounds, energies around $4.6$ meV should be avoided. As alternatives, energies of $3.7$ meV (the BeO filter edge) or $5.0$ meV (the Be-filter edge) can be strongly recommended. Our findings explain why, historically, best difraction results on stripes in cuprate superconductors was found for exactly these energies. Although the multiple scattering explanation of spurious signals was noted early by Tranquada {\em et al.}, we have here presented a comprehensive investigation of this effect that we now believe to be fully understood.

We emphasize the need for applying these considerations for the planning of neutron diffraction experiments on similar systems, in order to minimize contamination by spurious signals.

\section*{Acknowledgements}
We thank Jean-Claude Grivel and Maria Retuerto for valuable assistance with the preparation of the single crystal samples.

We are grateful to Philippe Bourges for assisting us with a related test experiment at the LLB and for bringing the idea of multiple scattering to our attention.

We thank the students from the University of Copenhagen Neutron Scattering Course as well as Linda Udby, Ursula B. Hansen, and Sonja Holm-Dahlin for participating in the initial measurements.

We thank Anne Bartholdy and Henrik Jacobsen for assisting in the neutron scattering characterization of the structural phase transition of the crystals.

This work was supported by the Danish Agency of Science and Technology through DANSCATT. Ana Elena \cb{T}u\cb{t}ueanu was supported by a Ph.D. grant from the Institute Laue-Langevin.

The neutron scattering data were obtained at the SINQ neutron source, Paul Scherrer Institute, Switzerland.


\nocite{*} 
\bibliographystyle{ios1}           
\bibliography{bibliography}        

\end{document}